\documentclass[10pt,a4paper]{article}  
\pdfoutput=1

\usepackage[english]{babel}

\usepackage{amsmath}
\usepackage{amssymb}
\usepackage{graphicx} 

\usepackage{feynmp} 
\usepackage{cancel}

\usepackage[colorlinks=true,pdfstartview=FitH]{hyperref} 

\usepackage{collref}

\newcommand{\Z}{\mathbb{Z}} 
\newcommand{\abs}[1]{|#1|} 

\title{$\Z_3$ Scalar Singlet Dark Matter}
\date{}
\author{
Genevi\`{e}ve B\'{e}langer,$^a$
Kristjan Kannike,$^{b,c}$ \\
Alexander Pukhov$^d$
and Martti Raidal$^{c,e}$
}

\begin{document}

\maketitle

\begin{center}
$^a${LAPTH, Univ. de Savoie,\\ CNRS, B.P.110, F-74941 Annecy-le-Vieux Cedex, France} 
\\
$^b${Scuola Normale Superiore and INFN, \\
Piazza dei Cavalieri 7, 56126 Pisa, Italy}
\\
$^c${National Institute of Chemical Physics and Biophysics, \\
R\"{a}vala 10, Tallinn, Estonia}
\\
$^d${Skobeltsyn Institute of Nuclear Physics,\\ Moscow State University, Moscow 119992, Russia} 
\\
$^e${Institute of Physics, University of Tartu,  Estonia}

\end{center}

\begin{abstract}
We consider the minimal scalar singlet dark matter stabilised by a $\Z_3$ symmetry. Due to the cubic term in the scalar potential, semi-annihilations, besides annihilations, contribute to the dark matter relic density. Unlike in the $\Z_2$ case, the dark matter spin independent direct detection cross section is no more linked to the annihilation cross section.
 We study the extrema of the potential and show that a too large cubic term would break the $\Z_3$ symmetry spontaneously,
 implying  a lower bound on the direct detection cross section, and allowing the whole parameter space to be tested by XENON1T. 
In a small region of the parameter space the model can avoid the instability of the standard model vacuum up to the unification scale.
If the semi-annihilations are large, however, new physics will be needed at TeV scale because the model becomes non-perturbative. 
The singlet dark matter mass cannot be lower than $53.8$~GeV due to the constraint from Higgs boson decay into dark matter.
\end{abstract}

\begin{fmffile}{Z3_scalar_singlet}
\fmfset{arrow_len}{2mm}

\section{Introduction}
\label{sec:introduction}

The most popular candidates for dark matter (DM) of the Universe are weakly interacting massive particles (WIMPs).
WIMPs have been searched for in direct as well as in indirect detection experiments, without success so far. 
Therefore the properties of DM are not known yet.

One popular class of WIMPs is scalar singlet DM \cite{Silveira:1985rk,McDonald:1993ex,Burgess:2000yq,Barger:2007im} (see also \cite{Barger:2008jx,Barger:2010yn,Guo:2010hq,Biswas:2011td,Drozd:2011aa}). 
Because of the recent discovery of the Higgs boson \cite{Chatrchyan:2012fk,Aad:2012gk},
we know that fundamental scalars do exist in Nature. Scalars that have the same gauge and $B-L$ quantum numbers as the standard model (SM)
elementary fermions, quarks and leptons, can be embedded in $SO(10)$ and are among the most natural DM candidates~\cite{Kadastik:2009dj}. 
In this case a theoretically well motivated connection between 
the DM, ordinary matter and non-vanishing neutrino masses is realised via grand unification~\cite{Kadastik:2009cu}. The singlet DM could be connected to electroweak (EW) baryogenesis \cite{Cline:2012hg}. 
In the SM with the $125~\text{GeV}$ Higgs boson, $H$, the vacuum becomes unstable at some scale before the unification scale \cite{EliasMiro:2011aa,Holthausen:2011aa,Xing:2011aa}.
Scalar singlet DM, $S$, coupled to the Higgs boson via the $\abs{S}^2 \abs{H}^2$ term,
can make the theory consistent up to the unification scale \cite{Gonderinger:2009jp,Kadastik:2011aa,Chen:2012fa,Cheung:2012nb,Gonderinger:2012rd,EliasMiro:2012ay,Lebedev:2012zw}. Thus the scalar sector may play an important r\^{o}le both in particle physics and in cosmology.

The real singlet scalar  DM model, that is the simplest DM model, is also very predictive.
Basically the relic density constraint determines the value of the direct detection cross section for each DM mass. The most stringent existing constraint on DM spin-independent scattering cross section with nuclei
obtained recently by XENON100 \cite{Aprile:2012nq} starts to probe its physical parameter space, as will be discussed in section 6.
XENON1T \cite{Aprile:2012zx} should be able to either rule out the entire scenario or find a DM signal. These  results provide an incentive to study the phenomenology of  a generalised scenario of singlet scalar DM. 

Although the real scalar singlet is the simplest candidate for dark matter, the only choice of stabilising symmetry for a real particle is a $\Z_2$ parity. 
To consider $\Z_3$ --- or $\Z_N$, in general --- only makes sense for a complex field, of which the simplest case is the complex scalar singlet. 
Although na\"ively this extension of the model may look marginal, the DM phenomenology is modified in a substantial way.
The $\Z_3$ singlet DM model we consider is \emph{the} simplest model to have semi-annihilations \cite{Hambye:2008bq,Hambye:2009fg,Arina:2009uq,DEramo:2010ep} and the DM relic abundance predictions are modified.\footnote{See the predictions of the  dark matter model \cite{Belanger:2012vp} in the limit where the doublet is heavy. }  As a consequence, the model predictions for the DM abundance and 
for the spin-independent direct detection cross section are not in one-to-one correspondence as in the case of  the $\Z_2$ model.
Phenomenologically this implies that the present DM direct detection experiments are not able to test the singlet scalar DM scenario conclusively. 

In spite of the fact that the complex scalar singlet with $\Z_3$ symmetry is arguably the simplest extension of the real singlet model,
the model in its most minimal form has not been studied in detail in the literature. Complex singlet scalar and $\Z_3$ symmetry
have been considered in the context of a model of neutrino mass generation~\cite{Ma:2007gq}, but its DM phenomenology was not studied there. Similar DM phenomenology occurs also in a DM model based on $D_3$ symmetry~\cite{Adulpravitchai:2011ei}, but this model is more complicated than the one presented here.

The aim of this work is to formulate and to perform a detailed study of the minimal scalar singlet DM model based on $\Z_3$ symmetry.
We first study the scalar potential of the model and derive constraints on its parameters from the requirements of vacuum stability and perturbativity. We find the extrema of the potential and show that the cubic $\mu_3$ term cannot be too large even if we allow for metastability of the SM vacuum.
We then implement the model in micrOMEGAs \cite{Belanger:2006is,Belanger:2008sj,Belanger:2010gh} and study its predictions for the DM relic abundance and for the spin-independent 
direct detection cross section. We find that predictions for the latter may be substantially reduced compared to the $\Z_2$ scalar DM
model but possess a lower bound because the $\Z_3$ symmetric SM vacuum must be the global minimum. We study renormalisation effects of the potential and find that
large semi-annihilation effects require the new physics scale to be as low as TeV, possibly associated with compositeness of dark matter \cite{Frigerio:2012uc}. 
We conclude that  the model is verifiable in future direct detection
experiments as XENON1T.

The paper is organised as follows. In section~\ref{sec:model} we formulate the minimal scalar DM model based on $\Z_3$.
In section~\ref{sec:vacuum} we study the properties of its vacuum.
In section~\ref{sec:rge} we study running of the model parameters due to renormalisation group.
In section~\ref{sec:relic} we calculate the predictions of the model for DM relic abundance and for direct detection experiments.
We discuss our results in section~\ref{sec:conclusions}.

\section{$\Z_3$ Scalar Singlet Model}
\label{sec:model}

In addition to the Higgs doublet $H$, the scalar sector contains the complex singlet $S$. The most general renormalisable scalar potential of $H$ and $S$, invariant under the $\Z_3$ transformation $H \to 1$, $S \to e^{i 2 \pi/3} S$, is
\begin{equation}	
\begin{split}
  V_{\Z_3} &= \mu_{H}^{2} \abs{H}^{2} + \lambda_{H} \abs{H}^{4} 
  + \mu_{S}^{2} |S|^{2} + \lambda_{S} |S|^{4} \\
  &+ \lambda_{SH} |S|^{2} |H|^{2} + \frac{\mu_3}{2} (S^{3} + S^{\dagger 3}),
\end{split}
\label{eq:V:Z:3:singlet}
\end{equation}
where $\mu_H^2 <0$. Without loss of generality, we can take $\mu_3$ to be real, since its phase can be absorbed in the phase of the singlet $S$. 
Also note that because the potential is invariant under simultaneously changing $\mu_{3} \to -\mu_{3}$, $S \to -S$,  
physics cannot depend on the sign of $\mu_{3}$, and it suffices to consider $\mu_{3} \geqslant 0$.
The potential \eqref{eq:V:Z:3:singlet} is the only possible potential with the given field
contents that is invariant under the $\Z_3$ group. We can always choose $H$ to transform trivially under $\Z_3$. 
The alternative transformation for the singlet, $S \to e^{i 4 \pi/3} S$, gives the same potential.
In the study of the parameter space, we choose $M_h^2$, $M_S^2$, $\mu_3$, $\lambda_S$,  $\lambda_{SH}$ 
and $v$ as free parameters. We fix the Higgs mass to $M_h = 125.5~\text{GeV}$ \cite{Giardino:2012dp} and the Higgs VEV to $v = 246.22~\text{GeV}$. The other parameters are then defined by
\begin{equation}
\begin{split}
  \mu_H^{2} &= -\frac{M_h^2}{2}, \\
  \lambda_H &= \frac{1}{2} \frac{M_h^2}{v^2}, \\ 
  \mu_S^2 &= M_S^2 - \lambda_{SH} \frac{v^2}{2}.
\end{split}
\label{eq:parameters}
\end{equation}

The model is perturbative \cite{Lerner:2009xg} if $\abs{\lambda_S} \leqslant \pi$ and $\abs{\lambda_{SH}} \leqslant 4 \pi$. The unitarity conditions  are weaker.

\section{Vacuum Stability \& Extrema of the Potential}
\label{sec:vacuum}

The scalar potential \eqref{eq:V:Z:3:singlet} is bounded below if the quartic interactions satisfy the vacuum stability conditions
\begin{equation}
\begin{split}
  \lambda_H &> 0, \\ 
  \lambda_S &> 0, \\
  2 \sqrt{\lambda_H \lambda_S} + \lambda_{SH} &> 0.
\end{split}
\label{eq:vac:stab}
\end{equation}

If the conditions \eqref{eq:vac:stab} are fulfilled, the scalar potential possesses a finite global minimum. 
To study the stationary points, it is convenient to use $\abs{H}^2 = h^2/2$ and write the singlet in polar coordinates as $S = s e^{i\phi_S}$.
The equations for stationary points, obtained by setting the partial derivatives of the potential with respect to $h$, $s$ and $\phi_S$ to zero, are given by
\begin{equation}
\begin{split}
  0 &= h \left[ M_h^2 (h^2 - v^2) + 2 \lambda_{SH} v^2 s^2 \right], \\
  0 &= s \left[ 4 \lambda_S s^2 +\lambda_{SH} (h^2 - v^2) + 2 M_S^2 
  + 3 \mu_3 s \cos {3 \phi_S} \right], \\
  0 &= s \mu_3 \sin 3 \phi_S. 
  \label{eq:stationary:points}
\end{split}
\end{equation}

Because we have choosen $\mu_3\ge 0$, we have $\cos{3\phi_S}=-1$ in local minima of potential with $s \ne 0$. This gives threefold degenerate vacua with $\phi_S = \pi/3, -\pi/3, -\pi$ that are related by $\Z_3$ transformations.

The Eqs.~\eqref{eq:stationary:points} are  reduced to quadratic equations.
The stationary points can be classified by their symmetries. 
The stationary points are
\begin{enumerate}
  \item $(\text{EW},\Z_3)$ \quad Unbroken EW and $\Z_{3}$ symmetry, $v_h = v_s = 0$,
  \begin{equation} 
    V_{\text{EW},\Z_3}=0
  \end{equation}   
  \item $(\cancel{\text{EW}},\Z_3)$ \quad Standard Model vacuum $v_h^2 = v^2$, $v_s = 0$ which is invariant under 
the $\Z_3$ symmetry,
\begin{equation} 
  V_{\cancel{\text{EW}},\Z_3}=-\frac{M_h^2 v^2}{8}
  \label{physVac} 
\end{equation}   
  \item $(\text{EW},\cancel{\Z_3})$ \quad Two triplets of vacua with unbroken EW symmetry and broken $\Z_{3}$; 
these solutions exist under the condition 
  \begin{equation}
    D_{\text{EW},\cancel{\Z_3}}= 9 \mu_{3}^2 
    - 16 \lambda_{S} (2 M_S^2 - \lambda_{SH} v^2) \geqslant 0,
  \label{eq:Z:3:breaking:cond}
  \end{equation}
  and read
\begin{equation}
\begin{split}
  v_s &= \frac{3\mu_3 \pm \sqrt{D_{\text{EW},\cancel{\Z_3}}}}{8\lambda_S}>0, \\
  v_h &= 0
 \end{split} 
\label{sol_EW_noZ3}
\end{equation}
  \item $(\cancel{\text{EW}},\cancel{\Z_3})$ \quad Two sextuplets of vacua where both the EW symmetry and $\Z_{3}$ are broken; they exist only if
\begin{equation}
   D_{\cancel{\text{EW}},\cancel{\Z_3}}=  9 M_h^2 \mu_3^2 - 16 M_S^2
  (2 \lambda_S M_h^2 - \lambda_{SH}^2 v^2) \geqslant 0,
   \label{eq:Z:3:EW:breaking:cond}
 \end{equation}
and read
  \begin{equation}
 \begin{split}
  v_s &= \frac{1}{4} \, \frac{3 M_h^2 \mu_3 \pm M_h \sqrt{   D_{\cancel{\text{EW}},\cancel{\Z_3}}}}{2 \lambda_S M_h^2 - \lambda_{SH}^2 v^2}>0 ,\\
  v_h^2 &= v^2\left(1-\frac{2\lambda_{SH} v_s^2}{M_h^2}\right)>0.
  \end{split}
  \label{sol_noEW_noZ3}
 \end{equation}

\end{enumerate}

We demand that the SM vacuum $(\cancel{\text{EW}},\Z_3)$ be the global minimum. The EW symmetry has to be broken, but the completely symmetric $(\text{EW},\Z_3)$ vacuum lies always above the physical one and thus is not dangerous. On the other hand, if the $\Z_3$ symmetry were broken, the singlet would be unstable and could not be dark matter. The degenerate vacua with different values of ${\phi_S}$ would raise the danger of cosmological domain walls \cite{Vilenkin:1984ib}. Therefore the potential energies of the vacua with broken $\Z_3$ have to be compared with \eqref{physVac}. Note that these solutions appear and can be below the SM vacuum if $\mu_3$ is large enough. 

\begin{figure}[t]
\centering
  \includegraphics[width=0.95\textwidth]{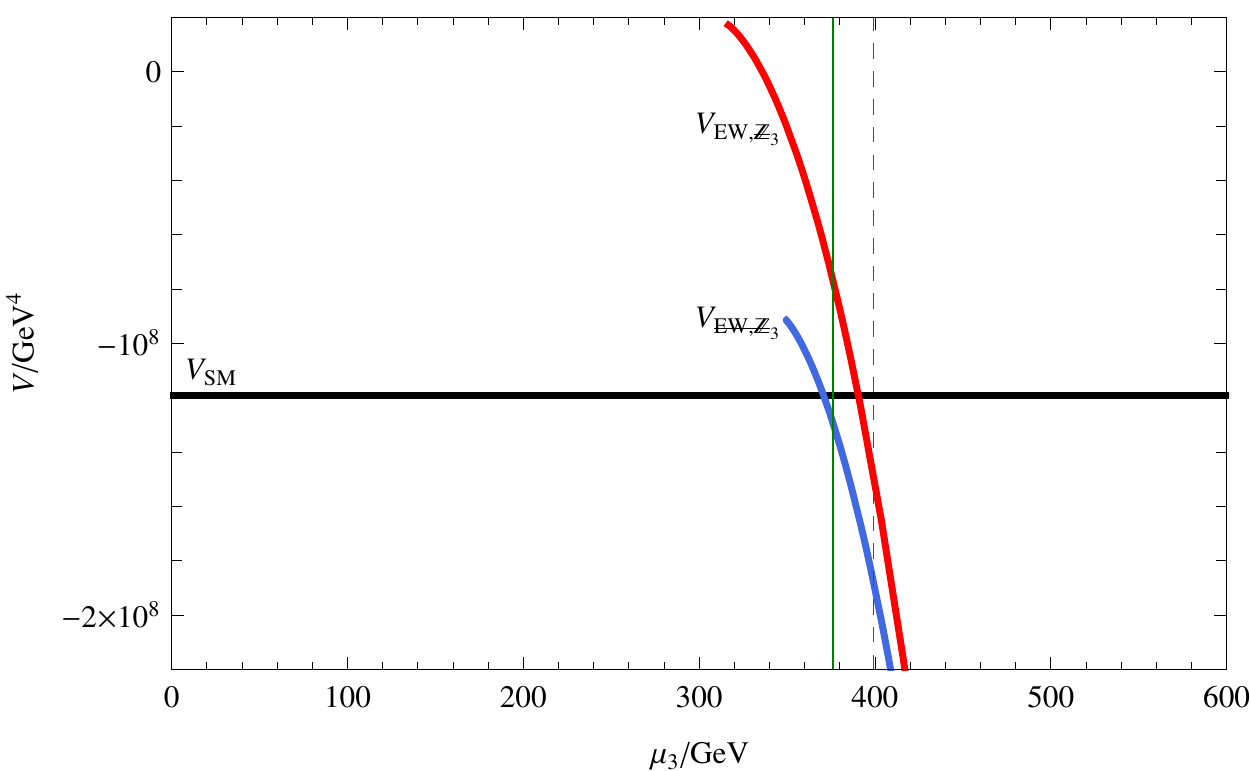}
\caption{Depencence of the energy of the stationary points on $\mu_3$. The values of other parameters are $M_S = 150~\text{GeV}$, $\lambda_S = \pi/2$, $\lambda_{SH} = 0.15$.  The black line is the potential of SM vacuum $(\cancel{\text{EW}},\Z_3)$, the red line is $V_{\text{EW},\cancel{\Z_3}}$, and the blue line is $V_{\cancel{\text{EW}},\cancel{\Z_3}}$. The vertical green line shows $\mu_3 \approx 2 \sqrt{\lambda_S} M_S$ where the energies of vacua become equal; the green dashed line shows $\max \mu_3$ allowed if our vacuum is metastable.
}
\label{fig:V:vs:muSii}
\end{figure}

This requirement gives a bound on $\mu_3$. We introduce a dimensionless parameter $\delta$ to  parameterise the energy difference of vacua \cite{Adams:1993zs}. Then the \emph{maximal} allowed value of the cubic parameter $\mu_3$ is approximately equal to
\begin{equation}
  \max \mu_3 \approx 2 \sqrt{2} \sqrt{\frac{\lambda_S}{\delta}} M_S
  \label{eq:max:mu''_S:vs:M_S}
\end{equation}
at given $\lambda_S$ and $M_S$, if $\abs{\lambda_{SH}}$ is small (as seen in section \ref{sec:relic} below, this is always true for realistic points with correct relic density). For $\delta = 2$, the $\Z_3$-breaking minimum $(\cancel{\text{EW}},\cancel{\Z_3})$ is approximately degenerate with the SM minimum and \eqref{eq:max:mu''_S:vs:M_S} gives the absolute stability bound.

 As $\mu_3$ grows larger, the potential energies of the $\Z_3$-breaking extrema rapidly descend below the value of the SM minimum. However, the bound \eqref{eq:max:mu''_S:vs:M_S} on $\mu_3$ could be relaxed with $\delta < 2$, if the SM vacuum were not the global minimum, but a metastable local minimum with a longer half-life than the lifetime of the universe. For $\Z_2$ singlet scalar DM, metastability bounds were calculated in \cite{Profumo:2010kp}. 

To estimate the metastability bound, we take $\lambda_{SH} \approx 0$ and use the results for a general quartic potential of a single scalar field  \cite{Adams:1993zs} which in our case is $s$ (the potential is already minimised with respect to the singlet phase $\phi_S	$). The decay probability per unit time per unit volume \cite{Coleman:1977py,Callan:1977pt} is given by
\begin{equation}
  \frac{\Gamma}{V} = K e^{-S_{\text{E}}},
\end{equation}
where $K$ is a determinantal factor and $S_{\text{E}}$ is the four-dimensional Euclidean action. In our case, $S_{\text{E}}$ is given by \cite{Adams:1993zs}
\begin{equation}
  S_\text{E} = \frac{\pi^2}{3 \lambda_S} \frac{1}{(2 - \delta)^3} 
  \left( \alpha_1 \delta + \alpha_2 \delta^2 + \alpha_3 \delta^3 \right),
\label{eq:SE}
\end{equation}
where $\alpha_1 = 13.832$, $\alpha_2 = -10.819$, $\alpha_3 = 2.0765$. The value of $K$ has very little influence on the allowed value of $\delta$ and can be approximated by the barrier height between two vacua. To ensure metastability, the probability of bubble nucleation in the past four-volume $1/H_0^4$, where $H_0 = 9.51 \times 10^{-42}$~GeV is the Hubble constant, must be 
\begin{equation}
  \frac{1}{H_0^4} \frac{\Gamma}{V} \leqslant 1.
  \label{eq:bubble:probability}
\end{equation}
The minimal allowed value of $\delta$ can be found by solving the equality in \eqref{eq:bubble:probability}.

The behaviour of minima of the potential is illustrated in figure~\ref{fig:V:vs:muSii} for a typical parameter set given by $M_S = 150~\text{GeV}$, $\lambda_S = \pi/2$, $\lambda_{SH} = 0.15$. The potential energies of $(\text{EW},\cancel{\Z_3})$ (red line) and $(\cancel{\text{EW}},\cancel{\Z_3})$ (blue line) fall below $V_{\text{SM}}$ after $\mu_3$ surpasses the bound \eqref{eq:max:mu''_S:vs:M_S} (green line). The bound on $\mu_3$ from metastability is shown with a dashed green line.

\section{Renormalisation Group Running}
\label{sec:rge}

Because of the running of couplings,  the vacuum may not be absolutely stable up to the Grand Unified Theory (GUT) scale, furthermore the model may become non-perturbative. We will study the influence of the running couplings on perturbativity and vacuum stability and see in which region the model can be valid up to the GUT scale.

The largest uncertainty on the vacuum stability bound arises from the top quark mass: the recent NLO \cite{EliasMiro:2011aa,Holthausen:2011aa,Xing:2011aa} and NNLO analyses \cite{Bezrukov:2012sa,Degrassi:2012ry} that use the top pole mass disfavour SM vacuum stability, though a couple of analyses at NNLO \cite{Masina:2012tz,Alekhin:2012py} that determine the running top mass from total top pair production cross section, allow vacuum stability up to Planck scale. In the model considered here, vacuum stability at the GUT scale fares better than in the SM, indeed even if the top contribution cannot guarantee vacuum stability, a large  Higgs-singlet coupling $\lambda_{SH}$ gives a positive contribution to the running of $\lambda_H$ \cite{Gonderinger:2009jp,Kadastik:2011aa,Chen:2012fa,Cheung:2012nb,Gonderinger:2012rd,EliasMiro:2012ay,Lebedev:2012zw} and solves the issue.

In the next section we will see that the semi-annihilation contribution to the DM relic density increases with $\mu_3$. Large values for this parameter require a large $\lambda_S$, Eq.~\eqref{eq:max:mu''_S:vs:M_S}, which in turn implies that due to  renormalisation group equation (RGE)  running that the model becomes non-perturbative at a relatively low scale. The perturbativity bound therefore strongly constrains models with significant semi-annihilation.

\begin{figure}[tb]
\begin{center}
  \includegraphics[width=1\textwidth]{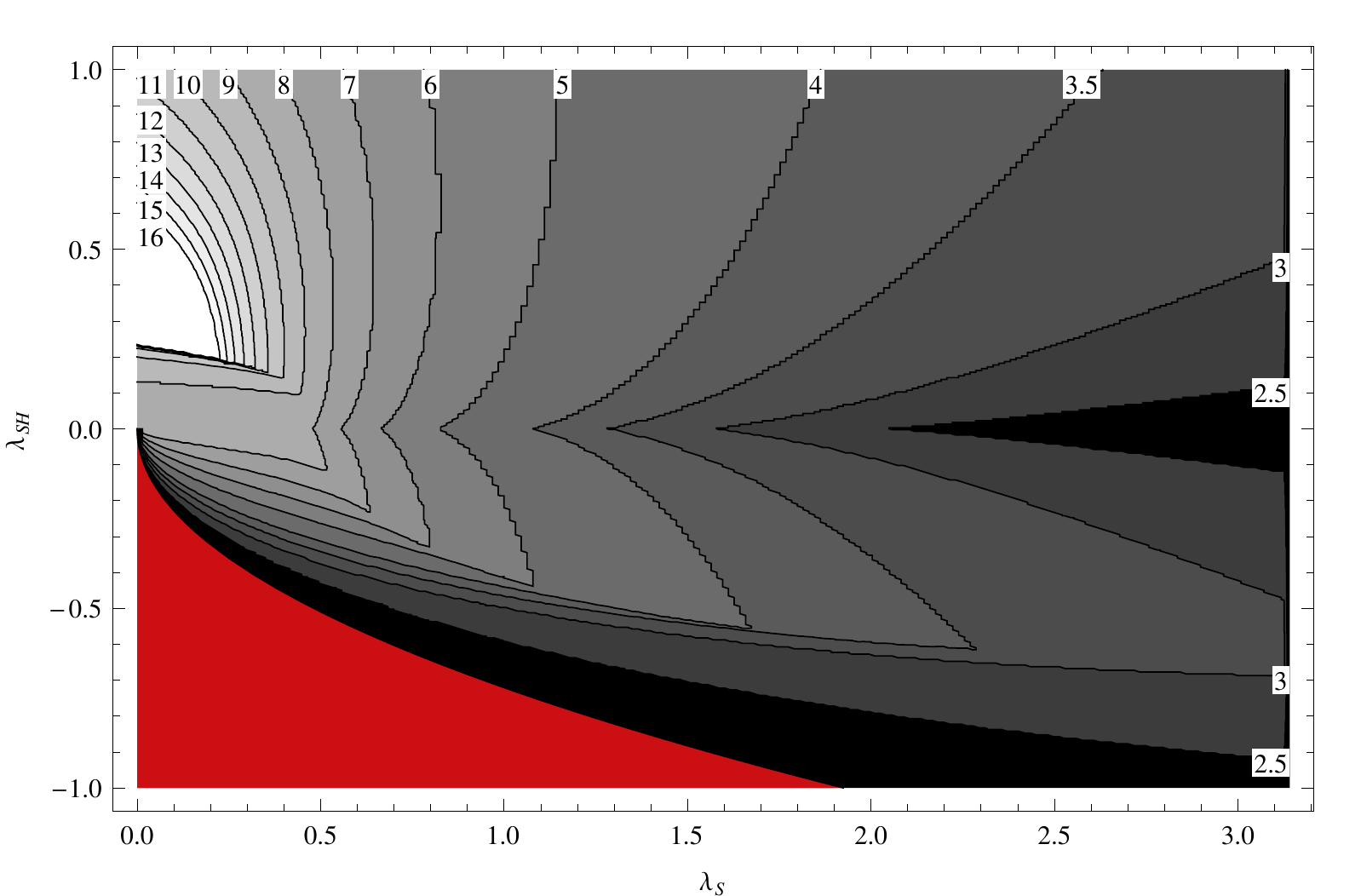}
\caption{Perturbativity and vacuum stability limits on the cutoff scale on the $\lambda_{SH}$ \textit{vs.} $\lambda_{S}$ plane. The contours show the logarithm of the renormalisation scale $\log_{10} \mu$ in GeV. In the white area the model is valid up to the GUT scale. The red area at bottom left is excluded by EW scale vacuum stability.
}
\label{fig:lambda:SH:vs:lambda:S}
\end{center}
\end{figure}

The values of the input parameters \cite{Beringer:1900zz} in the $\overline{\text{MS}}$ scheme are $\alpha_{\text{EM}}^{-1}(M_Z) = 127.944 \pm 0.014$, $\alpha_3 (M_Z) = 0.1196 \pm 0.0017$, $\sin^2 \theta_W (M_Z) = 0.23116 \pm 0.00012$, $M_Z = 91.1874 \pm 0.0021~\text{GeV}$. The running top quark mass is calculated \cite{Gray:1990yh} from the top quark pole mass \cite{Lancaster:2011wr} which is $M_t = 173.2 \pm 0.9~\text{GeV}$. The top quark is integrated in at its pole mass. In order to find the scale for realistic points, we integrate DM in at the scale given by the fit $M_{S} \approx (90.7 + 2070 \abs{\lambda_{SH}})$~GeV of $M_{S}$ as a function of $\abs{\lambda_{SH}}$ for the points in the WMAP $3 \sigma$ range, except the light mass range $M_{S} < 120$~GeV (see next Section).%
\footnote{For DM with light mass, this results in a slightly larger scale of loss of perturbativity, but the difference is negligible.}

We use the SM two-loop RGEs for the running of the gauge couplings and the top quark Yukawa coupling \cite{Ford:1992mv};\footnote{At one-loop level, the contribution of the singlet is zero; we neglect the two-loop contributions.} for the running of quartic scalar couplings we use the one-loop RGEs \cite{Kadastik:2009cu}
\begin{equation}
\begin{split}
  \kappa \beta_{\lambda_{H}} &= 24 \lambda_{H}^{2} - 3 (3 g^2 +  g^{\prime 2} - 4 y_{t}^2) \lambda_{H}
  \\ 
  &+ \frac{3}{8} (g^4 + 2 g^2 g^{\prime 2} + g^{\prime 4}) - 6 y_{t}^4 + \lambda_{SH}^2, \\
  \kappa \beta_{\lambda_{S}} &= 20 \lambda_{S}^{2} + 2 \lambda_{SH}^{2}, \\
  \kappa \beta_{\lambda_{SH}} &= 4 (3 \lambda_H + 2 \lambda_{S}) \lambda_{SH} 
  + 4 \lambda_{SH}^{2} 
  - \frac{3}{2} (3 g^{2} + g^{\prime 2} - 4 y_{t}^{2}) \lambda_{SH},
\end{split}
\label{eq:RGEs}
\end{equation}
where $\beta_{\lambda_i} \equiv d\lambda_i/d(\ln \mu)$, $\mu$ is the renormalisation scale, and $\kappa = 16 \pi^{2}$.

We take into account the $\overline{\text{MS}}$ corrections to the Higgs mass from the SM \cite{Casas:1994us} and from the singlet \cite{Gonderinger:2009jp}, and the corrections to $\lambda_H$ from the one-loop effective potential \cite{Casas:1994us,Espinosa:1995se}.

The results are shown in Figure~\ref{fig:lambda:SH:vs:lambda:S}. Because we consider the  value of $\lambda_{SH}$ correlated with $M_S$ for the points in the WMAP $3\sigma$ range, the scale at which the DM is integrated in and $\lambda_{SH}$ and $\lambda_S$ begin to run, is proportional to the distance from the horizontal axis. At large $\lambda_S$, the new physics scale is determined by loss of perturbativity, and can be just a few hundred GeV for $\lambda_S \lesssim \pi$.

For $\lambda_{SH} < 0$, it is the vacuum stability bound that sets the scale of validity. The red area in the lower left corner of Figure~\ref{fig:lambda:SH:vs:lambda:S} is already excluded by vacuum stability at the EW scale. As the Higgs self-coupling $\lambda_H$ runs to lower values, the last inequality in the vacuum stability conditions \eqref{eq:vac:stab} cannot be satisfied.

For small $\lambda_S$, the new physics scale is determined by $\lambda_{SH}$. For small and positive values of $\lambda_{SH}$, the Higgs self-coupling $\lambda_H$ behaves as in the SM and causes the vacuum to become unstable at about $10^{9}$~GeV. However, the RGE of $\lambda_H$ receives the positive contribution $\kappa \Delta \beta_{\lambda} = \lambda_{SH}^2$ from the singlet. For $\lambda_{SH} \gtrsim 0.2$ the vacuum will be stable up to the GUT scale. Increasing $\lambda_{SH}$ further, the loss of perturbativity brings the new physics scale slowly down again.

All in all, there is a small region in the $\lambda_{SH}$ \textit{vs.} $\lambda_S$ plane, where the model is perturbative up to the GUT scale, corresponding to $\lambda_S \lesssim 0.2$ and $0.2 \lesssim \lambda_{SH} \lesssim 0.5$.

\section{Relic Density \& Direct Detection}
\label{sec:relic}

\begin{figure}[tb]
\begin{center}
\begin{minipage}[b]{\linewidth}
\centering
\fmfframe(0,5)(4,2){
 \begin{fmfgraph*}(9,9)
 \fmfleft{i1,i2} 
 \fmfright{o1,o2}
 \fmf{dashes}{i1,v1,i2}
 \fmf{dashes}{o1,v1,o2} 
 \fmflabel{$S^*$}{i1}
 \fmflabel{$S$}{i2}
 \fmflabel{$h$}{o2}
 \fmflabel{$h$}{o1}
 \end{fmfgraph*}
 }
\fmfframe(0,5)(4,2){
 \begin{fmfgraph*}(18,9)
 \fmfleft{i1,i2} 
 \fmfright{o1,o2}
 \fmf{dashes}{i1,v1} 
 \fmf{dashes}{i2,v1}
 \fmf{dashes,label=$h$,label.side=left}{v1,v2}
 \fmf{dashes}{v2,o1} 
 \fmf{dashes}{v2,o2}
 \fmflabel{$S^*$}{i1}
 \fmflabel{$S$}{i2}
 \fmflabel{$h$}{o2}
 \fmflabel{$h$}{o1}
 \end{fmfgraph*}
 }
 \fmfframe(0,5)(4,2){
 \begin{fmfgraph*}(9,9)
 \fmfleft{i1,i2} 
 \fmfright{o1,o2}
 \fmf{dashes}{i2,v1,o2} 
 \fmf{dashes,label=$S$}{v1,v2}
 \fmf{dashes}{i1,v2,o1}
 \fmflabel{$S$}{i2}
 \fmflabel{$h$}{o2}
 \fmflabel{$S^*$}{i1}
 \fmflabel{$h$}{o1}
 \end{fmfgraph*}
 }
\fmfframe(0,5)(4,2){
 \begin{fmfgraph*}(18,9)
 \fmfleft{i1,i2} 
 \fmfright{o1,o2}
 \fmf{dashes}{i1,v1} 
 \fmf{dashes}{i2,v1}
 \fmf{dashes,label=$h$,label.side=left}{v1,v2}
 \fmf{boson}{v2,o1} 
 \fmf{boson}{v2,o2}
 \fmflabel{$S^*$}{i1}
 \fmflabel{$S$}{i2}
 \fmflabel{$V$}{o2}
 \fmflabel{$V$}{o1}
 \end{fmfgraph*}
 }
\fmfframe(0,5)(0,2){
 \begin{fmfgraph*}(18,9)
 \fmfleft{i1,i2} 
 \fmfright{o1,o2}
 \fmf{dashes}{i1,v1} 
 \fmf{dashes}{i2,v1}
 \fmf{dashes,label=$h$,label.side=left}{v1,v2}
 \fmf{fermion}{o2,v2,o1}
 \fmflabel{$S^*$}{i1}
 \fmflabel{$S$}{i2}
 \fmflabel{$\bar{f}$}{o2}
 \fmflabel{$f$}{o1}
 \end{fmfgraph*}
 }
 \\
 \vspace{3mm}
 (a)
\end{minipage}
\\
 \vspace{1mm}
 \begin{minipage}[b]{0.35\linewidth}
\centering
 \fmfframe(0,5)(0,2){
 \begin{fmfgraph*}(18,9)
 \fmfleft{i1,i2} 
 \fmfright{o1,o2}
 \fmf{dashes}{i1,v1} 
 \fmf{dashes}{i2,v1}
 \fmf{dashes,label=$S$,label.side=left}{v1,v2}
 \fmf{dashes}{v2,o1} 
 \fmf{dashes}{v2,o2}
 \fmflabel{$S$}{i1}
 \fmflabel{$S$}{i2}
 \fmflabel{$S^*$}{o2}
 \fmflabel{$h$}{o1}
 \end{fmfgraph*}
 }
\\
 \vspace{3mm}
(b)
\end{minipage}
\begin{minipage}[b]{0.35\linewidth}
\centering
 \fmfframe(0,5)(0,2){
 \begin{fmfgraph*}(9,9)
 \fmfleft{i1,i2} 
 \fmfright{o1,o2}
 \fmf{dashes}{i2,v1,o2} 
 \fmf{dashes,label=$h$}{v1,v2}
 \fmf{fermion}{i1,v2,o1}
 \fmflabel{$S$}{i2}
 \fmflabel{$S$}{o2}
 \fmflabel{$N$}{i1}
 \fmflabel{$N$}{o1}
 \end{fmfgraph*}
 }
 \\
 \vspace{3mm}
(c)
\end{minipage}
\end{center}
\caption{Feynman diagrams contributing to (a) annihilation and (b) semi-annihilation of dark matter; and (c) dark matter cross section with nucleons.
}
\label{fig:Feynman:diagrams}
\end{figure}
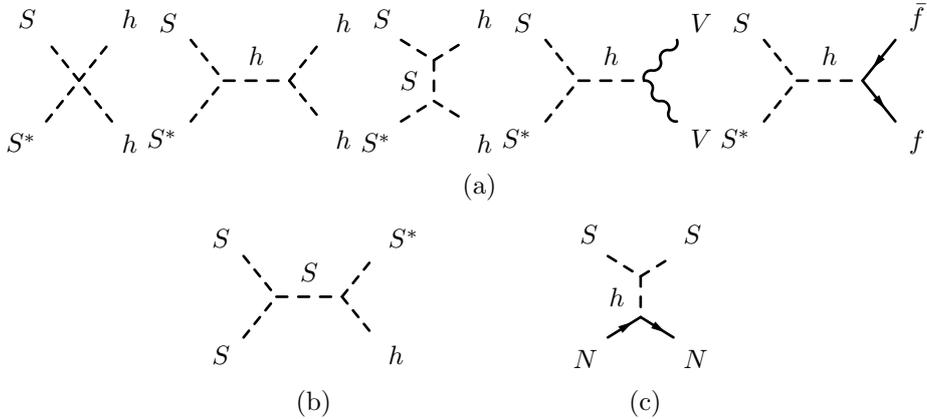

The presence of the semi-annihilation process can lower the annihilation cross section and thus the direct detection cross section with nucleons after taking into account the relic density constraint. Figure~\ref{fig:Feynman:diagrams} shows the Feynman diagrams that contribute to (a) annihilation and (b) semi-annihilation of dark matter, and (c) spin-independent interaction  with nucleons.

To compute the relic density we solve the Boltzmann equations with the micrOMEGAs package \cite{Belanger:2006is,Belanger:2008sj,Belanger:2010gh}. The equations for the number density, $n$, have been generalised to include semi-annihilation processes 
\begin{equation}
\frac{dn}{dt}=-v\sigma^{S S^* \rightarrow  XX}  \left(n^2-\overline{n}^2 \right) -\frac{1}{2} v\sigma^{SS\rightarrow
S^* h}  
\left(n^2-n \, \overline{n} \right) -3H n,
\end{equation} 
where $X$ is any SM particle. The treatment of the semi-annihilation term is  described in \cite{Belanger:2012vp} and  the fraction of semi-annihilation is defined as 
\begin{equation}
\alpha=\frac{1}{2} \frac{v\sigma^{SS\rightarrow S^* h}}{v\sigma^{SS^* \rightarrow XX}  +\frac{1}{2}v\sigma^{SS\rightarrow
S^* h}}.
\end{equation}
Note that $S S \to S^* h$ is the only  semi-annihilation process in this model.
In solving for the relic density, the annihilation processes into one real and one virtual gauge bosons \cite{Yaguna:2010hn} have been also taken into account: these can reduce the relic density by up to a factor of 3 in the region just below the $W$/$Z$ thresholds.\footnote{These processes will be available for any model in the next public version of micrOMEGAs.}

To study the parameter space, we scan over the free parameters in the ranges $1~\text{GeV} \leqslant M_S \leqslant 1000~\text{GeV}$, $0~\text{GeV} \leqslant \mu_3 \leqslant 4000~\text{GeV}$, $0 \leqslant \lambda_S \leqslant \pi$, $-4 \pi \leqslant \lambda_{SH} \leqslant 4 \pi$ with the uniform distribution. The upper bounds on $\lambda_S$ and $\lambda_{SH}$ come from perturbativity. 

We require each point to satisfy the vacuum stability conditions \eqref{eq:vac:stab} and the $\Z_3$ symmetric SM vacuum $(\cancel{\text{EW}},\Z_3)$ to be the global minimum to ensure that $S$ is stable. 

The WMAP survey bound on the relic density \cite{Larson:2010gs} is 
\begin{equation}
\Omega h^2 = 0.1009 \pm 0.0056.
\end{equation}
We choose the points in the WMAP $3\sigma$ range. 

\begin{figure}[tb]
\begin{center}
  \includegraphics[width=1\textwidth]{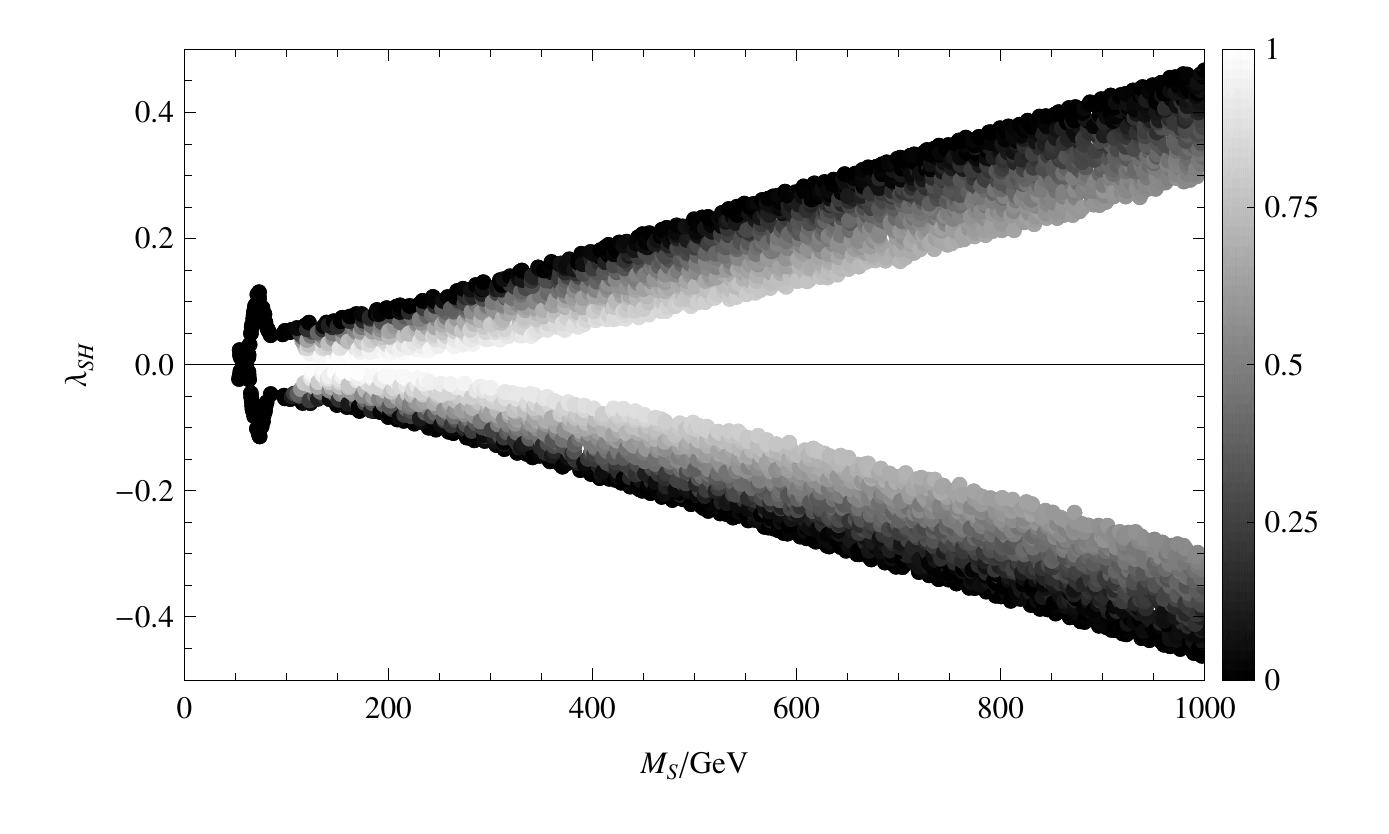}
\caption{
$\lambda_{SH}$ \textit{vs.} $M_S$ for the points in the WMAP $3 \sigma$ range satisfying $\mathrm{BR}_{\text{inv}} < 0.40$. Shading shows the fraction of semi-annihilation $\alpha$.}
\label{fig:lambdaSH:vs:MS}
\end{center}
\end{figure} 

For a heavy singlet, the dominant annihilation processes are into gauge bosons and Higgs pairs. In both cases the relic density is inversely proportional to $\lambda_{SH}^2/M_S^2$, and the WMAP constraint therefore selects a narrow band in the  $\lambda_{SH}$--$M_S$ plane, as seen in Figure~\ref{fig:lambdaSH:vs:MS}. In this figure the points are shaded by the fraction of semi-annihilation $\alpha$. For large values of $\alpha$, the contribution of annihilation processes to the relic density is suppressed so $\lambda_{SH}$ can be smaller. 

For kinematical reasons, semi-annihilation is only relevant for $M_S > M_h$. For a DM mass below the $W$ boson mass, annihilation is mainly into fermion pairs. Since the DM annihilation is suppressed by the Yukawa couplings of the fermions, a larger coupling $\lambda_{SH}$ is required, unless $M_S\approx M_h/2$ in which case the annihilation cross section is enhanced by a resonance effect and $\lambda_{SH}$ can be very small.

\begin{figure}[tb]
\begin{center}
  \includegraphics[width=1\textwidth]{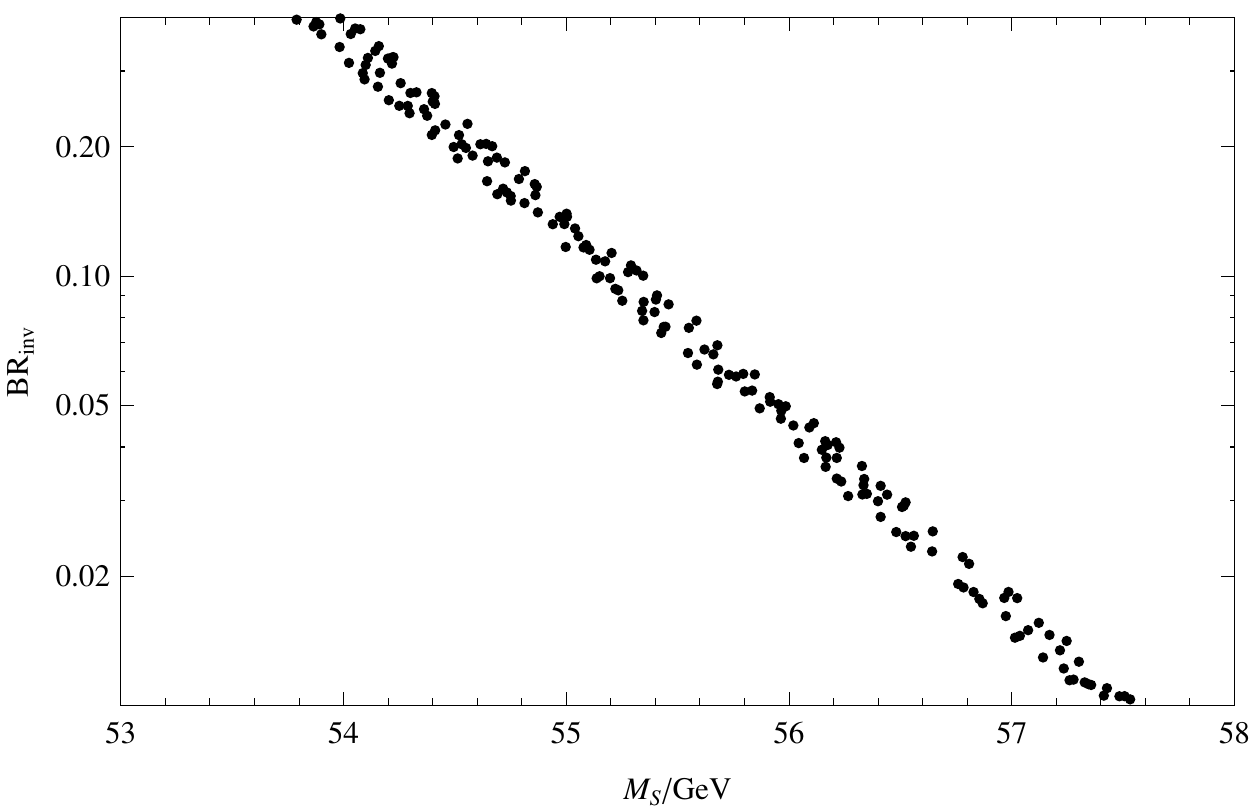}
\caption{Higgs boson invisible branching ratio $\mathrm{BR}_{\text{inv}}$ \textit{vs.} dark matter mass $M_S$. Mass range below $53.8$~GeV is excluded by $\mathrm{BR}_{\text{inv}}$.
}
\label{fig:BRinv:vs:M:S}
\end{center}
\end{figure}

Note that when the DM mass is below $M_h/2$, the Higgs can decay into two DM particles \cite{Raidal:2011xk,Mambrini:2011ik,Djouadi:2011aa}. This is in fact the dominant annihilation process of the Higgs leading to a mostly invisible decay  of the Higgs. This possibility has been severely restricted by the Higgs discovery at the LHC. Allowing for less than 40\% invisible width of the Higgs \cite{Giardino:2012dp,Giardino:2012ww,Espinosa:2012vu,Espinosa:2012im}, rules out most $M_S<M_h/2$ points. The only remaining points are those for which the invisible Higgs decay is phase-space suppressed. The Higgs invisible branching ratio $\mathrm{BR}_{\text{inv}}$ \textit{vs.} $M_S$ for these points is shown in Figure~\ref{fig:BRinv:vs:M:S}. Singlet masses below $53.8$~GeV are excluded by $\mathrm{BR}_{\text{inv}} > 0.40$.

\begin{figure}[t]
\centering
  \includegraphics[width=0.95\textwidth]{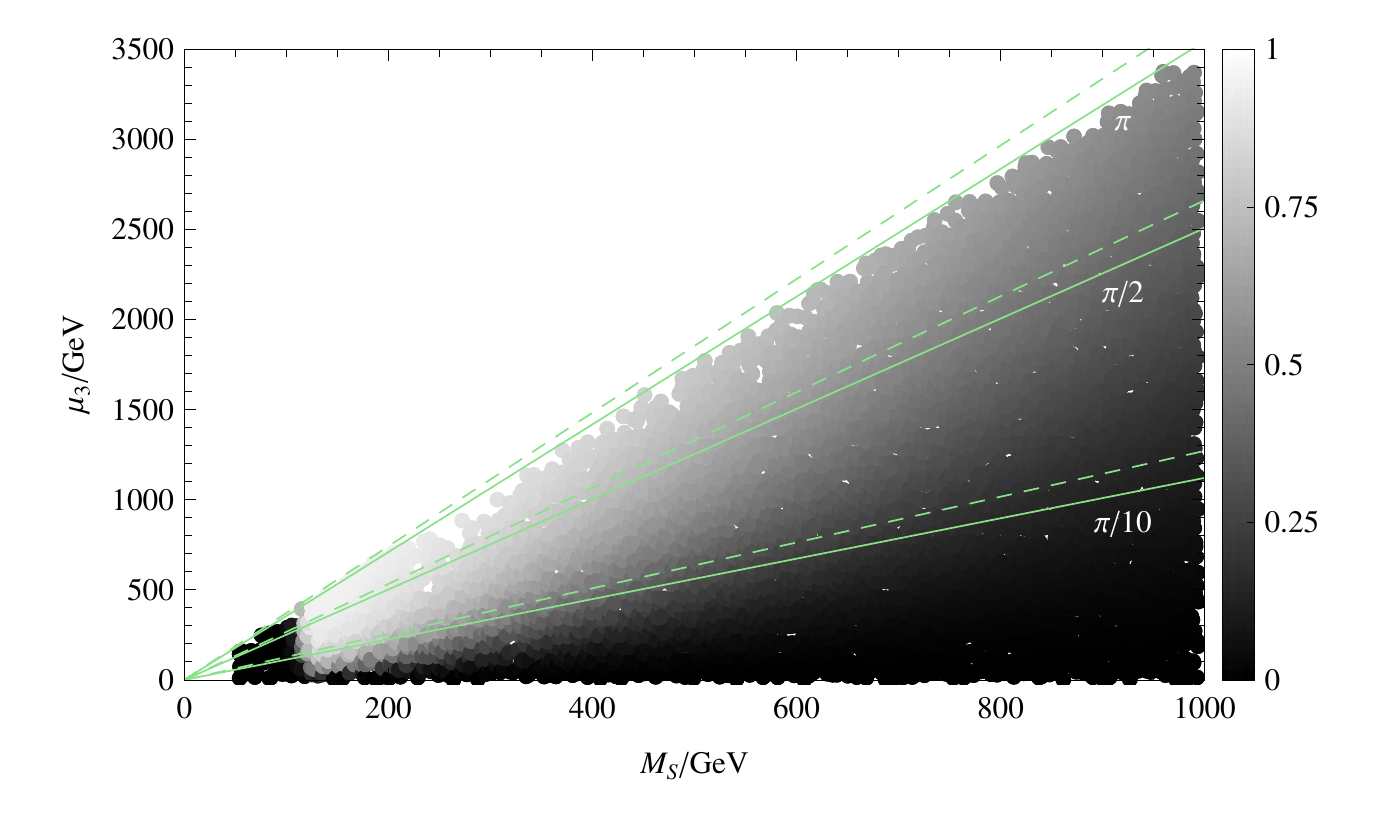}
\caption{
$\mu_3$ vs. $M_S$ for the points in the WMAP $3 \sigma$ range satisfying $\mathrm{BR}_{\text{inv}} < 0.40$. Green (dashed) lines show the bound \eqref{eq:max:mu''_S:vs:M_S} on $\mu_3$ from $\lambda_S$ from global stability (metastability). Shading shows the fraction of semi-annihilation $\alpha$. 
}
\label{fig:mu3:vs:MS}
\end{figure}

\begin{figure}[tb]
\begin{center}
  \includegraphics[width=1\textwidth]{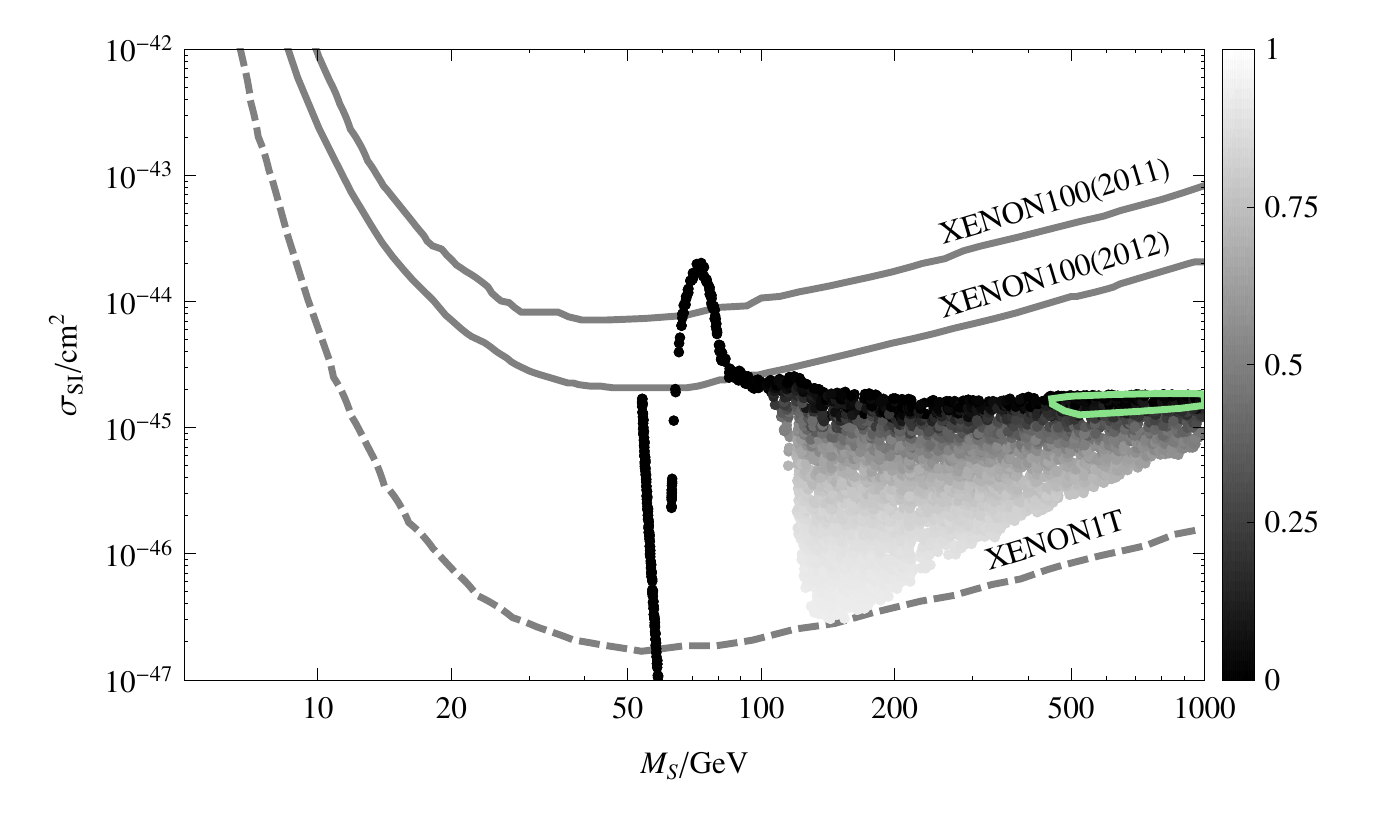}
\caption{Spin-independent direct detection cross section $\sigma_{\text{SI}}$ of $S$ with nucleons \textit{vs.}  dark matter mass $M_S$. The grey solid lines are the 90\% CL limits from the XENON100(2011) \cite{Aprile:2011hi} and the new XENON100(2012) results \cite{Aprile:2012nq}. The dashed grey line is the projected sensitivity of the XENON1T experiment \cite{Aprile:2012zx}. Shading shows the fraction of semi-annihilation $\alpha$. The parameter region encircled by green line is valid up to the GUT scale.
}
\label{fig:sigma:SI:vs:M:S}
\end{center}
\end{figure}

The points that do not satisfy the Higgs invisible decay constraint are not shown in Figures~\ref{fig:lambdaSH:vs:MS}, \ref{fig:mu3:vs:MS} and \ref{fig:sigma:SI:vs:M:S}.

Figure~\ref{fig:mu3:vs:MS} shows $\mu_3$ vs. $M_S$ for the points in the WMAP $3\sigma$ range and with $\mathrm{BR}_{\text{inv}} < 0.40$. The maximal $\mu_3$ at a given dark matter mass $M_S$ and the DM self-coupling $\lambda_S$ --- shown by green lines for global stability and dashed green lines for metastability --- is limited by the bound \eqref{eq:max:mu''_S:vs:M_S}. For $\lambda_{S} = \pi/10$, the parameter $\delta = 1.56$, for $\lambda_{S} = \pi/2$, $\delta = 1.77$, and for $\lambda_{S} = \pi$, $\delta = 1.83$. Since the Euclidean action $S_\text{E} \propto 1/\lambda_S$ \eqref{eq:SE}, the parameter $\delta$ can become small for small $\lambda_S$. However, because the bound \eqref{eq:max:mu''_S:vs:M_S} on $\mu_3$ is also proportional to $\sqrt{\lambda_S}$, the absolute size of the change is of the same order for all $\lambda_S$, though its \emph{relative} size is larger for smaller $\lambda_S$.

The points are shaded by the fraction of semi-annihilation $\alpha$. The semi-annihilation cross section goes as $\mu_3^2\lambda_{SH}^2/M_S^6$, hence it is largest for large values of $\mu_3$ and small values of $M_S$, corresponding to the lighter areas on the edges of the parameter space up to about $500~\text{GeV}$. In these areas the value of  $\abs{\lambda_{SH}}$ is smaller than expected because there semi-annihilation has a r\^{o}le in producing the correct relic density. 
The fact that the maximal value of $\mu_3$ is much smaller for low $M_S$ (Figure~\ref{fig:mu3:vs:MS}) somewhat tames the $M_S$ dependence in the relic density.

In Figure~\ref{fig:sigma:SI:vs:M:S} we display the spin-independent direct detection cross section $\sigma_{\text{SI}}$ \textit{vs.}  dark matter mass $M_S$ for the points in the WMAP $3\sigma$ range. Also shown are the XENON100 limits from 2011 \cite{Aprile:2011hi} and the new 2012 results \cite{Aprile:2012nq}, together with the projected sensitivity of the XENON1T experiment \cite{Aprile:2012zx}. The parameter region encircled by green line (with $M_S \gtrsim 450$~GeV) is valid up to the GUT scale. Since the spin-independent cross section is proportional to $(\lambda_{SH}/M_S)^2$, for large values of $M_S$ annihilation dominates the contributions to the relic density. If $\mu_3$ is small, the WMAP constraint basically imposes that $\sigma_{\text{SI}}\approx 2\times 10^{-45}~\mathrm{cm}^2$ for large masses. When semi-annihilation plays a r\^{o}le,  the coupling $\lambda_{SH}$ can be much smaller, and the spin independent cross section can be reduced by almost two orders of magnitude. The direct detection constraint  rules out the case where the singlet mass is below $M_W$ except for the very few points with $M_S < M_h/2$ that are still allowed because they correspond to  a small   invisible Higgs partial width. 

\section{Discussion \& Conclusions}
\label{sec:conclusions}

Phenomenologically, the simplest way to account for dark matter is to extend Standard Model with a scalar singlet. Indeed, real scalar singlet dark matter made stable by a $\Z_2$ symmetry is one of the best studied models. The $\Z_2$ model is very predictive, since the same singlet-Higgs coupling $\lambda_{SH}$ determines both the annihilation cross section and the spin-independent direct detection cross section $\sigma_{\text{SI}}$. Already the recent XENON100 results come close to discovering or excluding the $\Z_2$ model, and the model will be completely tested at the early XENON1T.

An equally valid choice of the stabilising symmetry is $\Z_3$. This change adds to the scalar potential the cubic $\mu_3$ term that produces a substantial change in the behaviour of the model. The $\mu_3 S^3$ term gives rise to the semi-annihilation process $S S \to S^* h$ that can dominate in determination of the relic density if $M_S > M_h$. Thus, the $\lambda_{SH}$ coupling can be smaller and the direct detection cross section can be lower than in the $\Z_2$ model. This will save scalar singlet dark matter even if early results from XENON1T will rule out the $\Z_2$ case.

Na\"ively it appears that $\lambda_{SH}$ could approach zero and $\mu_3$ become very large while keeping the product $\mu_3 \lambda_{SH}$ constant. The only process contributing to the relic density would be semi-annihilation; the annihilation and the direct detection cross section would be virtually nil. However, there is an upper bound \eqref{eq:max:mu''_S:vs:M_S} on $\mu_3$ that is proportional to $M_S$ and $\sqrt{\lambda_S}$. If the cubic term is too large, the $\Z_3$-symmetric SM vacuum is not the global minimum of the scalar potential. In principle, the SM vacuum could be metastable if the time of tunnelling is longer than the lifetime of the universe. Nevertheless allowing for metastability does not have a large impact on the parameter space. The effect is relatively small for small $M_S$ where semi-annihilation dominates the relic density. Therefore the change in the lowest value of $\sigma_{\text{SI}}$ which is obtained for $M_S$ slightly above $M_h$ in figure~\ref{fig:sigma:SI:vs:M:S} is at most 16\% for $\lambda_S = \pi$. At higher mass usual annihilation processes dominate in any case and the metastability bound has no impact.

We have implemented the model in micrOMEGAs and calculated the freeze-out relic density, taking into account both annihilation and semi-annihilation. We present analytic formulae for the extrema of the scalar potential. Demanding that the $\Z_3$-symmetric SM vacuum be the global minimum puts the upper bound \eqref{eq:max:mu''_S:vs:M_S} on $\mu_3$, and a lower bound on the fraction of semi-annihilation $\alpha$ for given $M_S$ and $\lambda_S$. The presence of semi-annihilation allows for smaller $\lambda_{SH}$ than annihilation only when $M_S > M_h$. Due to strong dependence of semi-annihilation on $M_S$, the direct detection cross section is lowest in the range from $M_h$ to about $200$~GeV. (Below $M_h$ the results are the same as for the $\Z_2$ complex singlet.) The model can be fully tested at XENON1T and other near future direct detection experiments even with the requirement that the $\Z_3$ symmetric SM vacuum be at least metastable.

If $M_S < M_h/2$, then Higgs boson can decay into dark matter. The unobserved Higgs boson invisible branching fraction $\text{BR}_{\text{inv}}$ excludes singlet masses $M_S \lesssim 53.8$~GeV. In the narrow range from $53.8$~GeV to $57.4$~GeV, the Higgs $\text{BR}_{\text{inv}}$ varies from $0.01$ to $0.4$ and could be measured by the LHC or a future linear collider.

The $\Z_3$-symmetric SM vacuum can be stable up to the GUT scale of $2 \times 10^{16}$~GeV if $\lambda_{SH} \gtrsim 0.2$. The positive contribution to the running of the Higgs self-coupling $\lambda_H$ counters the negative contribution from the top Yukawa. To be perturbative up to the GUT scale as well, one needs $\lambda_{SH} \lesssim 0.5$ and $\lambda_S \lesssim 0.2$. This corresponds to $M_S \gtrsim 450$~GeV and $\sigma_{\text{SI}} \approx (1.3 \ldots 1.8) \times 10^{-45}~\text{cm}^2$. If semi-annihilation is large, the model becomes unperturbative and new physics (new fermions or possibly a composite sector) has to come in at a few hundred GeV or at TeV scale.

\section*{Acknowledgements}

We thank Riccardo Barbieri for a suggestion. K.K. and M.R. were supported by the ESF grants 8090, 8499, 8943, MTT8, MTT59, MTT60, MJD140, by the recurrent financing SF0690030s09 project
and by the European Union through the European Regional Development Fund.
A.P. was supported by the Russian foundation for Basic Research, grants RFBR-10-02-01443-a and
RFBR-12-02-93108-CNRSL-a. The work of A.P. and G.B. was supported in part by the LIA-TCAP of CNRS.

\end{fmffile}

\bibliographystyle{JHEP}
\bibliography{Z_3_scalar_singlet}
\end{document}